\documentclass[aps,twocolumn,prd,showpacs,nofootinbib]{revtex4}
\usepackage{amsmath}
\usepackage{graphicx}
\usepackage{dcolumn}
\usepackage{bm}
\usepackage{amssymb}
\usepackage{latexsym}

\def\be{\begin{equation}}
\def\ee{\end{equation}}
\def\ba{\begin{eqnarray}}
\def\ea{\end{eqnarray}}

\begin{document}

\title{Galilean Islands in Eternally Inflating Background}

\author{Zhi-Guo Liu }
\author{Yun-Song Piao}

\affiliation{School of Physics, University of Chinese Academy of
Sciences, Beijing 100049, China}

\begin{abstract}

We show that the observational universe may emerge classically
from a de Sitter background with low energy scale. We find, after
calculating the curvature perturbation, that the resulting
scenario is actually a style of the eternal inflation scenario, in
which some regions will go through the slowly expanding Galilean
genesis phase with the rapidly increasing energy density and
become island universes, while other regions are still eternally
inflating, which will make the room for more island universes to
emerge.


\end{abstract}

\maketitle

\section{Introduction and Summary of Results}

During the eternal inflation
\cite{V1},\cite{L1},\cite{S1},\cite{GW1}, an infinite number of
universes will be spawned. It is generally thought that inside an
observational universe, a phase of the slow-roll inflation and
reheating is required, which will set the initial condition of the
``big bang" evolution, i.e. a homogeneous hot universe with the
scale-invariant primordial perturbation.

In principle, the slow-roll inflation should occur in a high
energy scale, which is required to insure that the amplitude of
primordial density perturbation is consistent with the
observations and as well as after the inflation, the reheating
temperature could be suitable for a hot ``big bang" evolution. In
this sense, it seems that the energy scale of the eternal
inflation should be enough high, or the spawning of observational
universe will be island-like, which is exponentially unfavored,
since it requires a large upward tunneling, e.g.
\cite{GV},\cite{DV},\cite{island},\cite{DKS}.

However, the observational universe might classically emerge from
a background with low energy scale, e.g. the emergent universe
scenario \cite{Ellis:2002we}. In the emergent universe scenario,
the universe originates from a static state in the infinite past,
when the universe emerges, or begins to deviate from this static
state, it is slowly expanding. In Ref.\cite{PZhou}, it was for the
first time observed that the scale-invariant curvature
perturbation might be adiabatically generated during the slow
expansion of primordial universe, also
\cite{JK},\cite{Piao1012},\cite{GKM}. Thus the initial conditions
of the ``big bang" evolution may be set after this slowly
expanding phase ends. During the slow expansion
\cite{PZhou},\cite{Piao1012}, the null energy condition is
violated, which might imply that the corresponding evolution
suffers from the ghost instability.

Recently, the application of Galileon \cite{NRT}, or its
nontrivial generalization, e.g.
\cite{Trodden},\cite{Vikman},\cite{KYY}, to the early universe has
acquired increasing attentions, e.g. see \cite{Gbounce} for
bouncing universe, and \cite{CNT},\cite{Liu:2011ns} for Galilean
genesis, in which the violation of null energy condition can be
implemented stably, there is not the ghost instability, and see
also earlier work about the ghost condensate
\cite{Buchbinder:2007ad} .
In Refs. \cite{Liu:2011ns}, it was showed by applying the
generalized Galileon that the scale invariant curvature
perturbation may be adiabatically generated in slowly expanding
Galilean genesis phase.

Here, inspired by \cite{Ellis:2002we}, we would like to ask a
significant question, whether and how the observational universe
may classically emerge from a de Sitter background with low energy
scale, and what about its scenario ?


We will show that the observational universe may emerge
classically from a de Sitter background with low energy scale. We
find, after calculating the curvature perturbation, that the
resulting scenario is actually a style of the eternal inflation
scenario, in which some regions will go through the slowly
expanding Galilean genesis phase and become island universes,
while other regions are still eternally inflating.

The outline of the paper is as follows. We firstly will introduce
a model, in which the observational universe may classically
emerge from a de Sitter background with low energy scale. The
background evolution of the model will be presented in Sec.II, and
the power spectrum of primordial perturbation will be calculated
in Sec.III. Here, though the energy scale of initial background is
highly low, since the universe slowly expands with rapidly
increasing energy density, the initial conditions of the hot ``big
bang" evolution may be set. In Sec.IV, we will illustrate the
resulting scenario, i.e. in an eternally inflating background,
some local regions may emerge classically and become island
universes. We argue that this scenario may be a viable design of
the early universe, which might help to improve the current
understanding to some issues relevant with the eternal inflation.

\section{The Background}

We will introduce a model, in which the observational universe may
classically emerge from a de Sitter background with low energy
scale. The method of building models is universal. Here, we will
begin with such a Galileon Lagrangian as
\be {\cal L}\sim -\,e^{4\varphi/{\cal M}}\,X+{1\over {\cal
M}^8}X^3-{1\over {\cal M}^7}X^2\Box\varphi-\Lambda, \label{L}\ee
where $X=\partial_\mu\varphi\partial^\mu \varphi /2$ and $\cal M$
is a constant with mass dimension, and $\Lambda$ sets up a de
Sitter background from which the observational universe emerges.





The evolution of background is determined by the field equation
\ba & &\left(-e^{4\varphi/ {\cal M}}+{15\over {\cal
M}^8}X^2+{24\over {\cal M}^7}H{\dot
\varphi}X\right){\ddot \varphi}\nonumber\\
&+& 3\left(-e^{4\varphi/ {\cal
M}}+{3\over {\cal M}^8}X^2\right)H{\dot \varphi}\nonumber\\
&+& \left(-{4\over {\cal M}}e^{4\varphi/ {\cal M}}+ {6{\dot
H}{\dot \varphi}^2\over {\cal M}^7}+{18{ H}^2{\dot \varphi}^2\over
{\cal M}^7}\right)X=0 \label{phi}\ea and the Friedmann equation
\be 3H^2M_P^2 =\, -\,e^{4\varphi/ {\cal M}}\,X+{5\over {\cal
M}^8}X^3 + {6\over {\cal M}^7}X {\dot \varphi}^3H+\Lambda
,\label{H}\ee where $M_P^2={1\over 8\pi G}$. When $H
> {{\dot \varphi}\over {\cal M}}$, the universe is in eternally
inflating regime, and the background is highly inhomogeneous, as
will be confirmed in Sec.IV. Thus initially we require \be H \ll
{{\dot \varphi}\over {\cal M}}.\label{condition}\ee The field
equation (\ref{phi}) approximately becomes \be \left(-e^{4\varphi/
{\cal M}}+{15\over {\cal M}^8}X^2\right){\ddot \varphi} -{4\over
{\cal M}}e^{4\varphi/ {\cal M}}X\simeq 0. \label{phiapp}\ee The
solution is \be e^{\varphi/{\cal M}}=\left({5\over
4}\right)^{1/4}{1\over {\cal M}(t_*-t)},\label{ephi}\ee \be {\dot
\varphi}={{\cal M}\over (t_*-t)}. \label{dphi}\ee
This implies $e^{4\varphi/ {\cal M}}\,X={5\over {\cal M}^8}X^3$.
Thus Eq.(\ref{H}) is simplified as, \ba H^2M_P^2  \simeq  {H\over
{\cal M}^2(t_*-t)^5} +\Lambda/3  =  {H M_P^2\over x^4
(t_*-t)}+\Lambda/3, \label{H11}\ea where we define \be x={\cal
M}^{1/2}M_P^{1/2}(t_*-t)\label{x}\ee for convenience. Initially
$|t|\gg |t_*|$, the first term in right-hand side of
Eq.(\ref{H11}) is negligible, which indicates that initially the
universe is in a de Sitter state. However, with the lapse of time,
we will have ${HM_P^2\over x^4 (t_*-t)}\simeq \Lambda$.
The corresponding time is \be t_{C}\sim  {-1\over H_0^{1/5}{\cal
M}^{2/5} M_P^{2/5}}= \left({H_0^{4/5}\over {\cal M}^{2/5}
M_P^{2/5}}\right)t_0, \label{tcri}\ee where
$H_0=\sqrt{\Lambda\over 3}/M_P=-1/t_0$. Thus after $t_C$, the
universe will deviate from the de Sitter state and begin to the
evolution of genesis. This is just the model which we require.


\subsection{The slowly expanding evolution}

We will detailed illuminate how the background evolves in this
model. During $t<t_{C}$, \be H= \left[1+{1\over
H_0x^4(t_*-t)}\right]^{1/2}H_0\simeq H_0 \ee is almost constant.
Thus \be a\sim e^{H_0 (t_C-t_0)},\ee which seems indicate that the
universe is exponentially expanding.
However, in term of Eq.(\ref{tcri}) and $H_0|t_0|=1$, we have \be
H_0(t_C-t_0) =\left(1-{H_0^{4/5}\over M_P^{2/5}{\cal
M}^{2/5}}\right) H_0|t_0| < 1, \label{Ht}\ee which implies that
the time that this phase lasts is shorter than one efold. Thus
during this period the universe is actually slowly expanding.

In certain sense, this phase is similar to the slow expansion
studied in Refs.\cite{JK},\cite{GKM}, also the slow contraction
\cite{KS1}. Here, $H\sim H_0+{1\over (t_*-t)^5}$, while in
Refs.\cite{JK},\cite{GKM}, $H\sim H_0+{1\over t_*-t}$.


When $t\simeq t_{C}$, the universe will deviate from the de Sitter
background and begin to the evolution of genesis. During $t>
t_{C}$, \be H\simeq {1\over x^4(t_*-t)} \label{H2}\ee is rapidly
increasing. Thus \be a \sim e^{\int Hdt} \sim Exp{\left({1\over
x^4}\right)}. \label{aa}\ee During this period, since $x\gg 1$,
the universe is still slowly expanding. However, different from
that during $t<t_C$, the energy density of universe during this
period will rapidly increase until the end of the slowly expanding
phase.

When $x\simeq 1$, the slow expansion ends. The definition
(\ref{x}) of $x$ gives \be t_e= {\cal O}(t_*)\simeq -{1\over
\sqrt{{\cal M}M_P}}. \label{te}\ee We assume that at $t_e$ the
reheating will happen and the available energy of field will be
rapidly released into the radiation. Hereafter, the local universe
will begin the evolution of hot ``big bang" model. Eq.(\ref{H2})
gives \be H_e\simeq {1\over x_e^4(t_*-t_e)} \simeq \sqrt{{\cal
M}M_P}. \ee Thus at this time, the energy density of Galileon
field is $ \rho_G\simeq M_P^3 {\cal M}\gg \Lambda$. In Sec.III.B,
we will see the observation requires ${\cal M}/M_P\sim 1/10^{10}$.
Thus this energy will be enough for the reheating of universe.

We plot the evolutions of $a$ and $H$ with respect to the time in
the inset panel of Fig.1, in which the parameters ${\cal M}=M_P$
and $\Lambda\sim M_P^4/10^{8}$ are taken, We have
$t_{C}/|t_e|\simeq -6$ from Eqs.(\ref{tcri}) and (\ref{te}), which
is consistent with Fig.1. Here, the values of the parameters used
are only to conveniently plotting the background evolution. In
principle, we could have a broader choice of the range of
parameters, e.g. $\Lambda$ is equal to or smaller than the value
of the current cosmological constant.

\subsection{The violation of null energy condition}

The statement of the null energy condition is equivalent to
$\epsilon>0$, where $\epsilon=-{{\dot H}\over H^2}$. Here, in term
of Eq.(\ref{H11}), we find ${\dot H}>0$ through the entire
evolution, which implies that during the slow expansion, the null
energy condition is violated all along.

Here, $H$ is determined by
Eq.(\ref{H11}), which is slightly complicated. However, for
different phases, we approximately have \ba |\epsilon |  =
\left|{{\dot H}\over H^2}\right| &\simeq & {{\cal M}M_P\over
H_{0}^2}/x^6,\,\,\,{\rm during} \,\,\,\,t<t_{C},\,\,\, \label{e1}\\
& & x^4,\,\,\,\,\,\,\,\,{\rm during}\,\,\,t_C<t<t_{e}, \label{e2}
\ea which implies that during $t<t_C$, $|\epsilon|$ is increasing
with the time, while it is decreasing during $t_C<t<t_e$. When
$t=t_C$, $|\epsilon|$ arrives at its maximal value \be
|\epsilon|\simeq {M_P^{2/5}{\cal M}^{2/5}\over H_0^{4/5}}\gg 1.
\ee
Physically, initially the universe is in a slow expanding phase
with $H\simeq H_0$ and ${\dot H}$ being gradually increased, thus
initially $|\epsilon|\ll 1$ and will become larger and larger,
while after $t>t_{C}$ the universe is in the slowly expanding
Galilean genesis phase with rapidly increasing $H$, thus
$|\epsilon|$ will be smaller and smaller. When $t_e\sim {\cal
O}(t_*)$, we have $|\epsilon|\simeq 1$, the genesis phase ends. We
plot the evolution of $\epsilon$ with respect to the time in
Fig.1.

Though during the slow expansion, the null energy condition is
violated all along, we will see that there is not the ghost
instability.

\begin{figure}[ht]
\includegraphics[scale=0.7,width=8.0cm]{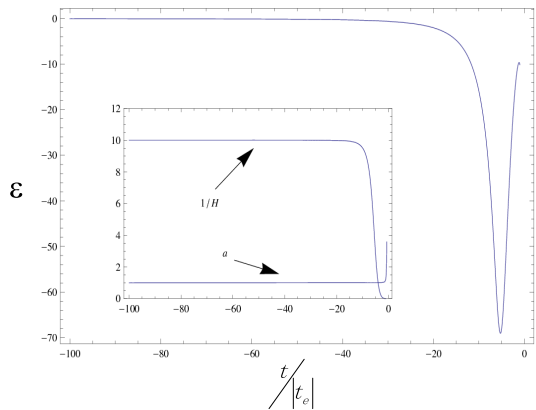}
\caption{The evolution of $\epsilon$ with respect to the time is
plotted, in which $|t_e|={1\over \sqrt{{\cal M}M_P}}$ is given by
Eq.(\ref{te}), and ${\cal M}=M_P$ and $\Lambda=M_P^4/10^{8}$ are
taken. The inset panel is the evolutions of $a$ and $1/H$, in
which initially $a_{ini}\simeq 1/H_{0}$, and for clarity $a$ has
been divided by $10^4$ and $1/H$ by $10^3$. Here,
$t_{C}/|t_e|\simeq -6$, which is given by Eq.(\ref{tcri}). We see
that
during $t<t_C$, the universe slowly expands with $H\simeq H_0$,
and during this period initially $|\epsilon|\ll 1$ and then will
gradually increase, up to $|\epsilon|\gg 1$, while during
$t_{C}<t<t_{e}$, the universe is still slowly expanding but with
rapidly increasing $H$ and decreasing $|\epsilon|$.  }
\label{fig:1}
\end{figure}



\section{The Perturbation}

We will study the curvature perturbation in this model. The
quadratic action of the curvature perturbation $\cal R$ is \be
S_{2}\sim \int d\eta d^3x {a^2Q_{\cal R}\over c_s^2}\left({{\cal
R}^\prime}^2-{c_s^2}(\partial {\cal R})^2\right), \label{S2}\ee
which has been calculated in the uniform field gauge in e.g.
Refs.\cite{Vikman},\cite{KYY}. Here, we have \cite{Liu:2011ns} \be
Q_{\cal R}= \left[{-e^{4\varphi/ {\cal M}}+{3X^2\over {\cal
M}^8}+{8X\over {\cal M}^7}({\ddot \varphi}+H{\dot
\varphi})-{8X^4\over {\cal M}^{14}M_P^2}\over \left({2{\dot
\varphi}X^2\over {\cal M}^7}-H\right)^2}\right]X, \label{QR}\ee
\be c_s^2={-e^{4\varphi/ {\cal M}}+{3X^2\over {\cal M}^8}+{8X\over
{\cal M}^7}({\ddot \varphi}+H{\dot \varphi})-{8X^4\over {\cal
M}^{14}M_P^2} \over -e^{4\varphi/ {\cal M}}+{15X^2\over {\cal
M}^8}+{12H{\dot \varphi}^3\over {\cal M}^7}+{12X^4\over {\cal
M}^{14}M_P^2}}, \ee both of which are determined by the evolution
of background.

\subsection{The evolutions without the ghost }

We will firstly investigate whether there is the ghost instability
around the corresponding backgrounds.


Before $t_e\sim {\cal O}(t_*)$, the universe is slowly expanding.
During this period the Galileon field is dominated, which
contributes the curvature perturbation $\cal R$. Here, $Q_{\cal
R}>0$ and $c_s^2>0$ are required for the avoidance of the ghost
instability. The evolution of field is determined by
Eqs.(\ref{ephi}) and (\ref{dphi}). We observe that in the
numerator of (\ref{QR}), the terms $\sim 1/x^6$ are dominated and
other terms are negligible, while in the denominator of
(\ref{QR}), during $t<t_C$, $H\simeq H_0$ is dominated, and during
$t_C<t<t_e$, $H$ is determined by Eq.(\ref{H2}). Thus we
approximately have \ba Q_{\cal R} & \simeq & {{\cal M}M_P^3\over
H_{0}^2}/x^6,\,\,\,{\rm during}\,\,\,t<t_{C}, \label{QQ}\\
&  & M_P^2 x^4,\,\,\,\,\,\,\,{\rm during}\,\,\,t_C<t<t_{e}.
\label{Q}\ea
The similar calculations give $c_{s}^2\sim 1.4$ during $t<t_e$.
Thus the background is ghost-free during the slow expansion. The
case is similar to that in ghost condensation mechanism, e.g. see
Ref.\cite{Buchbinder:2007ad} for the ekpyrotic universe. Here, it
is significant to notice $Q_{\cal R}\simeq M_P^2|\epsilon|$.


After $t_e\sim {\cal O}(t_*)$, the universe will be full of the
radiation, and begin the evolution of hot ``big bang" model.
During this period, Eq.(\ref{H}) become \be 3H^2M_{P}^2\simeq
\rho_{rad},\ee where $\rho_{rad}\sim
1/a^4$ is dominated. However, the Galileon field still exists,
though its energy density is negligible.

The perturbation of the Galileon field has been calculated in
Ref.\cite{Gcurvaton}. The field $\varphi$ is ghost-free requires
$Q_{\delta\varphi}>0$, in which \be Q_{\delta\varphi} =
-e^{4\varphi/ {\cal M}}+{3X^2\over {\cal M}^8}+{8X\over {\cal
M}^7}({\ddot \varphi}+H{\dot \varphi})-{8X^4\over {\cal
M}^{14}M_P^2}.\label{qdp}\ee During the reheating, the available
energy of Galileon field is rapidly released into the radiation,
it is reasonably imagined that after the reheating we have
${\ddot\varphi}\ll {\dot\varphi}^2/{\cal M}$ and $X\ll {\cal
M}^4$, i.e. $\dot \varphi\ll {\cal M}^2$. Thus the absence of the
ghost requires that \be e^{4\varphi/{\cal M}}<{X^2\over {\cal
M}^8} \label{con}\ee has to be satisfied, which equals to
$e^{\varphi/{\cal M}}<{{\dot \varphi}/{\cal M}^2}$. ${\dot
\varphi}/{\cal M}^2\ll 1$ implies that the condition (\ref{con})
is equivalent to $\varphi/ {\cal M}\ll -1$. The corresponding
value of $\varphi$ at the time $t_C$ can be obtained from
Eqs.(\ref{ephi}) and (\ref{tcri}), which is \be {\varphi_C\over
{\cal M}}\sim -\ln\left({{\cal M}^3\over M_P^2H_0}\right)^{1/5}\ll
-1. \label{phiC}\ee Thus if after the reheating the Galileon field
can be reset in a region from which it evolves initially, there
will be not the ghost instability during the hot ``big bang"
evolution.

\subsection{The power spectrum consistent with the observations}

After affirming that the background is ghost-free, we will
calculate the power spectrum of the curvature perturbation $\cal
R$ generated during the slow expansion. The equation of $\cal R$
is \be u_k^{\prime\prime} +\left(c^2_s k^2-{z^{\prime\prime}\over
z}\right) u_k = 0, \label{uk}\ee  after we define $u_k \equiv
z{\cal R}_k$, which can be derived from (\ref{S2}), where $'$ is
the derivative for the conformal time $\eta=\int dt/a\simeq t/a$
and $z^2\equiv 2a^2{Q_{\cal R}}/c_s^2$.

When $k^2\simeq z^{\prime\prime}/z$, the perturbation mode is
leaving the horizon, and hereafter it freezes out. Here, $a$ is
almost unchanged, which implies $z\sim \sqrt{Q}$. Thus with
Eqs.(\ref{QQ}) and (\ref{Q}), we can write $z^{\prime\prime}/z$
during different periods as \ba {z^{\prime\prime}\over z}
={\left({\sqrt Q}\right)^{\prime\prime}\over \sqrt{Q}} & \simeq &
{\left(\nu^2-{1\over 4}\right)/(\eta_*-\eta)^2},\ea where
$\nu^2={49/ 4}$ for $k<k_C$ and $\nu^2={9/ 4}$ for $k>k_C$, and
$k_C$ is the comoving wave number of the perturbation mode leaving
the horizon at $t=t_C$. Thus Eq.(\ref{uk}) is approximately a
Bessel equation with the $\cal R$ horizon \be 1/{ H}_{\cal
R}=a\sqrt{\left|{z\over z^{\prime\prime}}\right|}\sim t_*-t,
\label{HR}\ee which is right during $t<t_{C}$ and $t>t_{C}$.


When $k^2\gg z^{\prime\prime}/z$, i.e. the perturbation is deep
inside the $\cal R$ horizon, $u_k$ oscillates with a constant
amplitude. The quantization of $u_k$ is well-defined, which sets
its initial value, \be u_k\sim \frac{1}{\sqrt{2k}} \, e^{-ik\eta}.
\ee

When $k^2\ll z^{\prime\prime}/z$, the solution of $u_k$ is \ba u_k
& = & {\sqrt{\pi}\over 2}e^{i2\pi }
\sqrt{\eta_*-\eta}H_{7/2}^{(1)}\left(k(\eta_*-\eta)\right)
\nonumber\\ & \simeq & {4 e^{i3\pi/2} \Gamma({7\over 2})\over
\sqrt{2k} \Gamma({3\over 2})}/ \left(k(\eta_*-\eta)\right)^3 \ea
for $k<k_C$ and \ba u_k  & = & {\sqrt{\pi}\over 2}e^{i\pi }
\sqrt{\eta_*-\eta}H_{3/2}^{(1)}\left(k(\eta_*-\eta)\right)
\nonumber\\ & \simeq &{e^{i \pi /2}\over \sqrt{2k}} /
\left(k(\eta_*-\eta)\right)\ea for $k>k_C$, respectively. The
perturbation spectrum is \be {\cal P}_{\cal R} = {k^3\over
2\pi^2}\left|{u_k\over z}\right|^2,\ee which is \ba {\cal P}_{\cal
R} & \simeq & {{\cal M}^2H_0^2 a^4\over k^4} ={{\cal M}^2 \over
H_0^2} \left({k_{0}\over k}\right)^4,\,\,\,{\rm for} \,\,\,\,
k<k_C, \label{P1}\\
& \simeq & {{\cal M}\over M_P x^6},\,\,\,\,\,\,\,\,\,{\rm
for}\,\,\,\, k>k_C, \label{P2}\ea where $k_{0}={a H_{0}}$ is the
comoving wave number of the perturbation mode leaving the horizon
at certain time $t_0$. We see that in the region $k<k_C$, the
spectrum is highly red tilt, while in the region $k>k_{C}$, it is
scale-invariant, but its amplitude will increase $\sim 1/x^6$. We
also may write Eq.(\ref{P2}) as ${\cal P}_{\cal R}(k>k_{C})\simeq
|\epsilon| {H^2\over M_P^2}$ in term of Eqs.(\ref{H2}) and
(\ref{e2}).




The evolution of $\cal R$ outside the horizon is \ba
{\cal R} & \sim &
D_1\,\,\,\,\, is\,\,\,{{\rm constant}}\,\,\,{ {\rm mode}}\label{C}\\
&or &\, D_2\int {d\eta\over z^2}\,\,\,\,\, is\,\,\,{{\rm
changed}}\,\,\,{ {\rm mode}} , \label{D}\ea where the increase or
decay of the $D_2$ mode is dependent on the evolution of $z$. We
find that during $t<t_{C}$ the spectrum of $\cal R$ is dominated
by the constant mode, while during $t>t_{C}$ the spectrum of $\cal
R$ is dominated by the increasing mode, which is \be {\cal R}\sim
\int {d\eta\over Q_{\cal R}}\sim {1\over x^3}, \label{P6}\ee which
is consistent with Eq.(\ref{P2}).

Here, a significant thing is though during $t<t_{C}$ the amplitude
of perturbation having left the horizon is constant, it will
synchronously increase with the perturbation leaving the horizon
during $t>t_{C}$ after $t>t_C$. Noting that for the perturbation
mode outside the horizon, only is its amplitudes increasing, but
the tilt of the spectrum is not altered
\cite{Piao0901},\cite{ZLP}.

When $|\epsilon|\sim 1$ or $x\simeq 1$, the change of $a$ begins
to become not negligible, as has been mentioned in Sec.II. Thus
the increasing of the perturbation amplitude will come to a halt
shortly after $t_{e}\sim {\cal O}(t_*)$.
This implies that the power spectrum of $\cal R$ should be that
calculated around $t_{e}$. In term of Eq.(\ref{P2}), noting
$x_e\simeq 1$, we have
\ba {\cal P}_{\cal R} \sim {{\cal M}\over M_P}. \label{P3}\ea The
observations give ${\cal P}_{\cal R}^{1/2}\sim 1/10^{5}$, which
requires ${\cal M}/M_P\sim 1/10^{10}$. Thus in this model, the
parameter $\cal M$ may be fixed by the observations, there is only
a free parameter, i.e. $\Lambda$.

\begin{figure}[ht]
\includegraphics[scale=0.7,width=8.0cm]{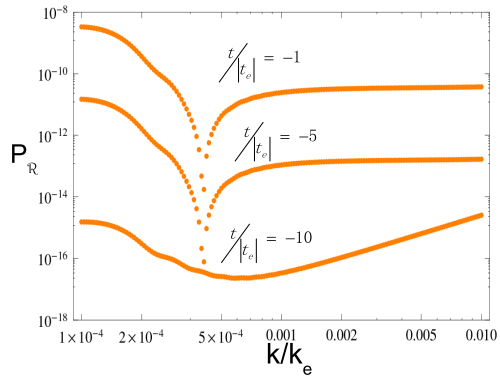}
\caption{ The perturbation spectra with respect to $k/k_e$ are
plotted at different times, in which $k$ is the comoving wave
number of the perturbation mode and $k_e$ corresponds to the mode
leaving the horizon at the end time $t_e$. We see that the
perturbation spectrum is almost scale-invariant for $k>k_C$, and
is $\sim 1/k^4$ for $k<k_C$. The amplitude of perturbation
spectrum increases with the time, but the tilt of the spectrum is
not altered. However, at $t/|t_e|=-10$ the spectrum is tilt for
$k>k_C$, the reason is that the corresponding perturbation modes
have still not left the horizon and the spectrum is set by the
initial condition.}
\end{figure}

\begin{figure}[ht]
\includegraphics[scale=0.7,width=8.0cm]{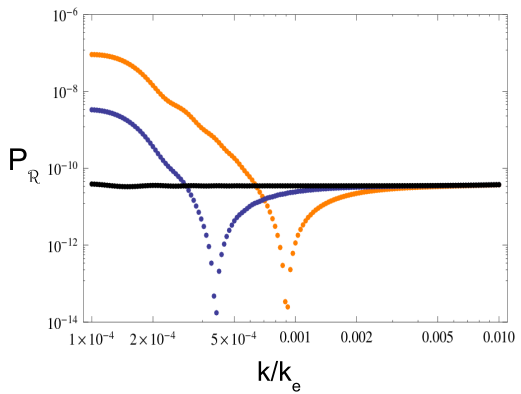}
\caption{ The perturbation spectra with respect to $k/k_e$ are
plotted for $\Lambda=0$ (black line), $\Lambda=M_P^4/10^{8}$ (blue
line) and $\Lambda=5\times M_P^4/10^{8}$ (orange line),
respectively, in which $k$ is the comoving wave number of the
perturbation mode and $k_e$ corresponds to the mode leaving the
horizon at the end time $t_e$. We see that for $\Lambda=0$, the
spectrum is scale-invariant on all scales, but for $\Lambda\neq
0$, the perturbation spectrum on scales larger than $1/k_C$ will
become highly red tilt $\sim 1/k^4$, and the larger $\Lambda$ is,
the larger $k_C$ is.}
\end{figure}

We numerically solved Eq.(\ref{uk}), and plotted the evolution of
the amplitude of the curvature perturbation in Fig.2 and the
resulting perturbation spectra for the different values of
$\Lambda$ in Fig.3. In Fig.2, we see that the perturbation
spectrum is almost scale invariant for $k>k_C$, and is $\sim
1/k^4$ for $k<k_C$, the amplitude of perturbation spectrum
increases with the time, but the shape of the spectrum is not
altered. Both Figs.2 and 3 are consistent with analytical results
in Eqs.(\ref{P1}) and (\ref{P2}).



\section{Galilean island in eternally inflating background}

We have showed that the observational universe may classically
emerge from a de Sitter background with low energy scale. We will
see what about the resulting scenario.

For $k<k_{C}$, the amplitude of the curvature perturbation is
determined by Eq.(\ref{P1}), which is highly red tilt. This result
implies that the amplitude of the perturbation will rapidly
increases with scale $1/k$. When $k/a=\sqrt{{\cal M}H_0}$, we have
${\cal P}_{\cal R} \sim 1$. We define the time when this
perturbation mode leaves the horizon as $t_{Eter}$. When the
corresponding perturbation mode leaves the horizon, i.e.
$k/a\simeq H_{\cal R}$, from Eq.(\ref{HR}), we have \be
t_*-t_{Eter}\simeq {1\over \sqrt{{\cal M}H_0}}. \ee Thus noting
$t_*=1/\sqrt{{\cal M}M_P}\ll 1/\sqrt{{\cal M}H_0}$, we have \be
t_{Eter}\sim -{1\over \sqrt{{\cal M}H_0}}\sim \sqrt{H_0\over {\cal
M}} t_0, \label{eternal}\ee and the corresponding field value
$\varphi_{Eter}$ can be derived from Eq.(\ref{ephi}), \be
\varphi_{Eter}={\cal M}\ln{1\over {\cal M}(t_*-t)}\sim -{\cal
M}\ln\left({{\cal M}\over H_0}\right). \ee Thus during
$t<t_{Eter}=\sqrt{H_0\over {\cal M}} t_0$, or equivalent
$\varphi<\varphi_{Eter}$, the energy density $\rho_\varphi$ of
local regions will be randomly walking. This in certain sense
implies that the global universe is actually in an eternal
inflating state, i.e. some regions have gone or are going through
the Galilean genesis phase, but other regions are still in
inflationary regime, the inflation never completely ends.

The classical evolutions of local universes begin only after
$t_{Eter}$, not $t_0$ as used in Eq.(\ref{Ht}). However, since
$t_0<t_{Eter}<t_C$, the result is not affected.


In principle, the eternal inflation will occur in any region of
space where the amplitude of the density perturbation $\sim 1$. In
slow-roll inflation model with single normal field, ${\cal
P}_{\cal R}\sim 1$ implies that the perturbation $\delta \phi$ of
inflaton $\phi$ is the same order as its classical rolling
$\Delta\phi\sim {{\dot \phi}/H}$ in unite of Hubble time, since
${\cal R}={H\over {\dot\phi}}\delta\phi\sim
{\delta\phi/\Delta\phi}$.

Here, we will see that ${\cal P}_{\cal R}\sim 1$ similarly means
that the perturbation of field is the same order as its classical
rolling. The perturbation of the Galileon field has been
calculated in Ref.\cite{Gcurvaton}.
When $k^2\ll z_{\delta\varphi}^{\prime\prime}/z_{\delta\varphi}$,
we have \be \delta\varphi_k\simeq {\Gamma({5\over 2})e^{i\pi}\over
 a \sqrt{2k Q_{\delta\varphi}}\Gamma({3\over
2})} /\left(k(\eta_*-\eta)\right)^2, \ee where
$z_{\delta\varphi}=a\sqrt{Q_{\delta\varphi}}$ and \be
Q_{\delta\varphi}\simeq {M_P^2\over {\cal M}^2}/ x^4,\ee which is
given by Eq.(\ref{qdp}). Thus the average square of the amplitude
of field fluctuations is \ba <\delta\varphi^2_k> & = & {1\over
8\pi^3}\int^{aH_{\delta\varphi}}_{aH_{\delta\varphi}/e}
\left|\delta\varphi_k\right|^2 d^3k \simeq \int^{aH_{\delta\varphi}}_{aH_{\delta \varphi}/e} {a^2{\cal M}^4\over k^3} dk \nonumber\\
& \sim & {{\cal M}^4\over H_{\cal R}^2}, \ea where \be 1/{
H}_{\delta\varphi}=a\left|{z_{\delta\varphi}\over
z_{\delta\varphi}^{\prime\prime}}\right|^{1/2}\sim t_*-t\ee
corresponds to the horizon of $\delta\varphi$, which means that
for the perturbation being leaving the horizon, we have $k/a\simeq
H_{\delta\varphi}$. We actually have $H_{\cal R}\simeq
H_{\delta\varphi}$. The classical rolling of Galileon field is
approximately \be \Delta\varphi={{\dot\varphi}/ H_0}\sim {{\cal
M}H_{\cal R}\over H_0}, \ee where $\dot \varphi$ is given by
Eq.(\ref{dphi}). Thus for $k<\sqrt{{\cal M}H_0}$, we have \be
{\sqrt{<\delta\varphi^2_k>}\over \Delta\varphi}\sim {{\cal
M}H_0\over H_{\cal R}^2}>1. \ee Thus during $t<t_{Eter}$, or
equivalent $\varphi<\varphi_{Eter}$, the perturbation of field is
larger than its classical rolling, which implies that the field is
randomly jumping. Thus in some local regions of the global
universe, the field will jumped to the regime of
$\varphi>\varphi_{Eter}$, and the local universe will go through
the evolution of slowly expanding Galilean genesis, while in other
regions the field will again jump back and is still in random
jumping. This result again indicates that the global universe is
actually in an eternally inflating regime.



When $t\sim t_{Eter}$, we have \be {{\dot \varphi}/{\cal
M}}=\sqrt{{\cal M}\over H_0} H_0\gg H_0, \ee which insures the
rationality of (\ref{condition}).

\begin{figure}[ht]
\includegraphics[scale=0.7,width=8.0cm]{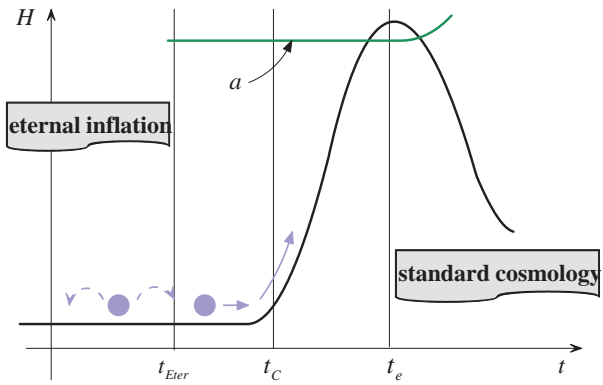}
\caption{ The genesis of the Galilean island in eternally
inflating background. The black line is the evolution of the
Hubble parameter $H$ of island universe with respect to the time,
while the green line is that of $a$. In this scenario, the global
universe is eternally inflating. During the eternal inflation, in
some local regions, the field will jump to the regime of
$\varphi>\varphi_{Eter}$ and then will classically evolve with the
initial time $t\simeq t_{Eter}$. During $t_{Eter}<t<t_{C}$, the
local universe is in a slowly expanding phase with $H\simeq H_0$,
while during $t_{C}<t<t_{e}$, the local universe is in a slowly
expanding genesis phase with rapidly increasing $H$. When $t=t_e$,
the genesis phase of local universe ends, and the universe
reheats. Hereafter, it will enter into the evolution of standard
cosmology. Here, we call this local thermalized universe as island
universe. During different periods after $t_{Eter}$, the values of
$H$ is summarized in Tab.I.}
\end{figure}


\begingroup
\begin{table}
    \label{spectrum}
    \begin{tabular}{|c|c|c|c|c|}
       \hline \hline  {} & \multicolumn{2}{|c|}{$H$ } &
       \multicolumn{2}{|c|}{$|\epsilon|$ } \\
      \hline \hline
      $t=t_{Eter}$  & \multicolumn{2}{|c|}{$ H_0$} &
       \multicolumn{2}{|c|}{$\ll 1$}
       \\
      \hline  \hline
      $t_{Eter}<t<t_C$  & \multicolumn{2}{|c|}{$H_0$} &
       \multicolumn{2}{|c|}{
       ${1 \over H_{0}^2{\cal M}^2M_P^2(t_*-t)^6}$
       }\\
      \hline  \hline
      $t=t_C$  & \multicolumn{2}{|c|}{$\gtrsim H_0$} &
       \multicolumn{2}{|c|}{${M_P^{2/5}{\cal M}^{2/5}\over H_0^{4/5}}\gg 1$ }\\
      \hline \hline
      $t_C<t<t_e$  & \multicolumn{2}{|c|}{${1\over M_P^2{\cal M}^2(t_*-t)^5}$} &
       \multicolumn{2}{|c|}{$M_P^2{\cal M}^2(t_*-t)^4$ }\\
      \hline \hline
      $t=t_e$  & \multicolumn{2}{|c|}{$\sqrt{{\cal M}M_P}\gg H_0$} &
       \multicolumn{2}{|c|}{$\sim 1$ }\\
      \hline \hline
    \end{tabular}
    \caption{
During different periods after $t_{Eter}$, the values of $H$ and
$|\epsilon|$ of island universe
    are summarized.
    Through the entire evolution, $a$ is almost unchanged and
    $\epsilon<0$.
    Though the null energy condition is violated,
    there is not the ghost instability, as has been confirmed in Sec.III.A.
     }
  \end{table}
  \endgroup

We summarized the resulting scenario in Fig.4. During the eternal
inflation, some regions of global universe will go through the
slowly expanding Galilean genesis phase and become island
universes, while other regions are still eternally inflating,
which will make the room for more island universes to emerge. The
island universes generally has the initial conditions
$a_{ini}\simeq 1/H_0$ and $H_{ini}\simeq H_0$, which is consistent
with Ref.\cite{FG}. During different periods after $t_{Eter}$, the
values of $H$ and $|\epsilon|$ are summarized in Tab.I. We see
that during $t_{Eter}<t<t_{C}$, the island universe is in a slowly
expanding phase with almost constant $H\simeq H_0$, while during
$t_{C}<t<t_{e}$, it is in the Galilean genesis phase with rapidly
increasing $H$ and the scale-invariant primordial perturbation may
be generated during this period. At $t=t_e$, the genesis phase of
island universe ends, the available energy of field will be
released into the radiation, and the universe reheats,
e.g.\cite{PZhang},\cite{Liu:2011ns}. Hereafter, in the
corresponding local region, or islands, the evolution of hot ``big
bang" model begins.


Here,
the method of building models is actually universal. The cubic
Galileon or the conformal Galileon \be {\cal L}\sim
-\,e^{2\varphi/{\cal M}}\,X+{1\over {\cal M}^4}X^2-{1\over {\cal
M}^3}X\Box\varphi \label{L1}\ee is apparently simpler. However,
with the conformal Galileon, the adiabatic perturbation during the
corresponding Galileon genesis is blue tilt \cite{CNT}, while with
(\ref{L}) the adiabatic perturbation is scale-invariant. Though
the resulting scenario is essentially not affected by this result,
since inside islands the obtaining of the scale invariance of
curvature perturbation may appeal to either the conversion of the
perturbations of other light scalar fields, e.g. conformal
mechanism \cite{Rubakov},\cite{HK}, or a period of inflation after
the emergence of Galilean islands, it may be imagined that the
introduction of extra field will make the building of model with
the conformal Galileon slightly complicated. In principle, the
implement with DBI genesis \cite{HJKM} is essentially same.
However, the relevant issues will be undoubtedly interesting for
investigating, while the simpler scenario showed here is a
starting point.

Here, $\Lambda$ is regarded as constant, which sets up a de Sitter
background from which the observational universe emerges. The
model with $\Lambda$ being a landscape of effective potential of
Galileon field $\varphi$ or other fields is certainly far
interesting. This issue will be studied in upcoming work.

In the discussions on the measure problem for eternally inflating
multiverse, e.g. \cite{V2},\cite{Bousso}, it is generally thought
that the emergence of island-like universe requires a large upward
tunneling, which is exponentially unfavored. In certain sense, our
scenario is a significant supplement to the phenomenology of the
eternal inflation, which might help to improve the current
understanding for the measure problem. The relevant issue will be
discussed elsewhere. The introduction of AdS bounce in eternally
inflating background has similar result \cite{Garriga:2012bc}, and
also \cite{Piao:2004me},\cite{Johnson:2011aa}.

\textbf{Acknowledgments} This work is supported in part by NSFC
under Grant No:11075205, 11222546, in part by the Scientific
Research Fund of GUCAS(NO:055101BM03), in part by National Basic
Research Program of China, No:2010CB832804.

\end{document}